\documentclass[prl,twocolumn,superscriptaddress,showpacs]{revtex4-1}

\usepackage{graphicx}

\begin{document}

\title{Common Crystalline and Magnetic Structure of superconducting $A_{2}$Fe$_{4}$Se$_5$ 
}

\author{F. Ye}
\affiliation{Neutron Scattering Science Division, Oak Ridge National Laboratory, Oak Ridge, Tennessee 37831, USA}
\author{S. Chi}
\affiliation{Neutron Scattering Science Division, Oak Ridge National Laboratory, Oak Ridge, Tennessee 37831, USA}
\author{Wei Bao}
\email{wbao@ruc.edu.cn}
\affiliation{Department of Physics, Renmin University of China, Beijing 100872, China}
\author{X. F. Wang}
\author{J. J. Ying}
\author{X. H. Chen}
\affiliation{Hefei National Laboratory for Physical Science at Microscale \\and Department of Physics, University of Science and Technology of China, Hefei, Anhui 230026, China}
\author{H. D. Wang}
\author{C. H. Dong}
\author{Minghu Fang}
\affiliation{Department of Physics, Zhejiang University, Hangzhou 310027, China}

\date{\today}

\begin{abstract}
Single crystal neutron diffraction studies on superconductors $A_{2}$Fe$_{4}$Se$_5$, where $A$ = Rb, Cs, (Tl,Rb) and (Tl,K) ($T_c\sim 30$ K), uncover the same Fe vacancy ordered crystal structure and the same block antiferromagnetic order as in K$_{2}$Fe$_{4}$Se$_5$. The Fe order-disorder transition occurs at $T_S= 500$-578 K, and the antiferromagnetic transition at $T_N= 471$-559 K with ordered magnetic moment $\sim$3.3 $\mu_B$/Fe at 10 K.
Thus, all recently discovered $A$ intercalated iron selenide superconductors share the common crystalline and magnetic structure, which are very different from previous families of Fe-based superconductors, and constitute a distinct new 245 family.  
\end{abstract}

\pacs{74.70.Xa,74.25.Ha,75.25.-j,75.30.-m}

\maketitle

The iron pnictide and chalcogenide superconductors, as exemplified by 
the $Re$FeAs(O,F) \cite{Kamihara2008,A033603,A033790,A042053}, (Ba,K)Fe$_2$As$_2$ \cite{A054630}, Li$_{1-x}$FeAs \cite{A064688} and Fe$_{1+x}$Se \cite{A072369} families of materials, share the Fe square lattice as the common 
structure feature, which determines electronic states at the Fermi surfaces responsible
for high-$T_c$ superconductivity \cite{A032740}.
The Fe$_{1+x}$(Se,Te) and Li$_{1-x}$FeAs have the same crystal structure \cite{A092058,A111613,A072228}, in which
the excess Fe and Li deficiency at the same crystallographic site serve  
to maintain close to +2 oxidation state of the Fe ions.
Recently, K$_{0.8}$Fe$_2$Se$_2$, regarded as K intercalated FeSe in the BaFe$_2$As$_2$ structure,
has been discovered as a superconductor with $T_c$ above 30 K \cite{C122924}.
Subsequent works replacing K by Rb or Cs \cite{C123637,C125525}
have led to more new superconductors of $T_c$ up to $\sim$32 K.
The heavy departure of Fe from the +2 valency implied by the nominal chemical formulas seem supported by angle-resolved photoemission (ARPES) measurements \cite{C125980}, which would upset theoretical
foundation of the prevailing mechanism of previously discovered Fe-based superconductors \cite{D014390,D014988,D016056,D023655}. 

\begin{figure}
\includegraphics[width=.94\columnwidth]{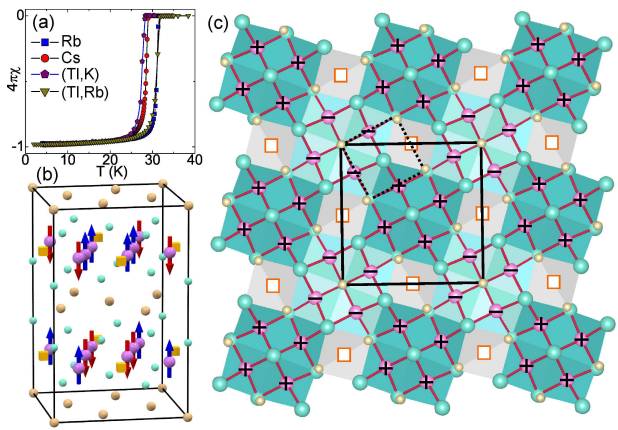}
\vskip -.2cm
\caption{(a) Magnetic susceptibility of $A_{2}$Fe$_{4}$Se$_5$ [$A$=Rb, Cs, (Tl,K) and (Tl,Rb)] showing the superconducting transition. (b) Common crystal and magnetic structure in the tetragonal $I4/m$ unit cell. The cyan sphere represents Se,  yellow sphere the intercalating $A$, the orange cube the Fe1 vacancy site, and the purple sphere the occupied Fe2 site with +(-) sign indicating up(down) magnetic moment along the $c$-axis. There are 4 formula units per cell. (c) The top Fe-Se layer.
The dark and light cyan
plaquettes emphasize the four-spin blocks of the up and down spins respectively, which forms a checkerboard nearest-neighbor antiferromagnetic pattern. From one Fe layer to another, magnetic moments are antiferromagnetically coupled.
The solid line denotes the $I4/m$ unit cell, one of the two twins, which breaks the high-temperature symmetry of the $I4/mmm$ unit cell (dashed line) at the Fe order-disorder transition at $T_S$.
}
\label{fig1}
\end{figure}

However, x-ray and neutron diffraction structure refinement studies have shown the correct compositions 
for the $A$=K or Cs intercalated iron selenide superconductors as close to $A_{0.8}$Fe$_{1.6}$Se$_2$  \cite{D014882,D020830}. Thus, the new superconductors are not heavily electron doped as previously
thought, and Fe has a formal oxidation state close to +2 as in the Fe$_{1+x}$(Se,Te) superconductors \cite{A092058,A111613}. The $\sim$20\% Fe vacancies in K$_{0.8}$Fe$_{1.6}$Se$_2$
are not randomly distributed in the BaFe$_2$As$_2$ structure below an order-disordered
transition at $T_S\approx 578$ K \cite{D020830}. Instead, they
form an ordered structure with the Fe1 site almost empty and the Fe2 site fully
occupied in the $\sqrt{5}$$\times$$\sqrt{5}$$\times$1 supercell as shown in Fig.~\ref{fig1}~(b)-(c) \cite{D020830}. 
At a slightly lower
temperature $T_N\approx 559$ K, magnetic moments carried by Fe ions at the Fe2 site develop a block checkerboard antiferromagnetic order.
The Fe vacancy and antiferromagnetic order also drastically alter the topology of the Fermi surface according to band structure calculations \cite{D021344,D022215}, which would have important physics ramifications and affect the interpretation of the ARPES \cite{C125980,D014556,D014923} and optic \cite{D010572} data.

Also reported are (Tl,K)Fe$_{y}$Se$_2$ and (Tl,Rb)Fe$_{1.72}$Se$_2$ superconductors of similarly high $T_c\sim 30$ K \cite{C125236,D010462}. What distinguishes these superconductors from the Tl-less ones is the appearance of the superconducting state at the quantum critical point of
an antiferromagnetic insulator \cite{C125236}. 
It is imperative then to find out whether the K, Cs or Rb superconductors and the (Tl,K) or (Tl,Rb) superconductors are two different types of new Fe selenide superconductors.
We report here single-crystal neutron diffraction investigation on the Rb, Cs, (Tl,K) and (Tl,Rb) superconductors from about 10 to 580 K that shows the same Fe vacancy and antiferromagnetic order as in the K$_{0.8}$Fe$_{1.6}$Se$_2$ superconductor, and the identification of a low $T_N$  in the 
previous bulk study \cite{C125236} is incorrect. Hence, superconductivity in all five $A_{0.8}$Fe$_{1.6}$Se$_2$ [$A$=K, Cs, Rb, (Tl,K) and (Tl,Rb)] superconductors develops in a common
crystalline and magnetic structure framework, which is distinctly different from the previous Fe square lattice-based families of iron superconductors \cite{Kamihara2008,A054630,A064688,A072369}. Per chemistry convention, $A_{0.8}$Fe$_{1.6}$Se$_2$ is better written as $A_{2}$Fe$_{4}$Se$_5$ 
for the new 245-family of superconductors.

Single crystals of Rb$_{2}$Fe$_{4}$Se$_{5}$ ($T_c\approx 32$ K) \cite{C125525},  Cs$_{2}$Fe$_{4}$Se$_{5}$ ($T_c\approx 29$ K) \cite{C125552}, (Tl,K)$_2$Fe$_4$Se$_5$ ($T_c\approx 28$ K) \cite{D022783} and (Tl,Rb)$_2$Fe$_4$Se$_5$ ($T_c\approx 32$ K) \cite{C125236} superconductors were grown using the Bridgeman method at USTC and ZU, and bulk superconductivity is indicated by nearly 100\% diamagnetic response reported in previous studies using samples grown with the same recipe\cite{C125552,C125525,C125236,D022783} and shown in Fig.~1(a).  
A small piece of Cs$_{2}$Fe$_{4}$Se$_{5}$ single crystal
from the same batch was used in the previous x-ray crystal structure refinement study at 295 K \cite{D014882}.
Single crystal neutron diffraction experiments were performed at the High Flux Isotope Reactor (HFIR) of the Oak Ridge National Laboratory. The Wide Angle Neutron Diffractometer (WAND) was used to
collect Bragg peaks in the ($h0\ell$), ($2h,h,\ell$) and ($3h,h,\ell$) reciprocal planes at various temperatures. Vertically focused Ge(113) monochromator was used to produce neutron beam of wavelength $\lambda=1.460 \AA$. The curved, one-dimensional $^3$He position-sensitive detector of 624 anodes covers 125$^o$ of the scattering angle. The samples were also investigated using the Four-Circle Diffractometer (FCD) with neutrons of $\lambda=1.536 \AA$ at selected temperatures. The order parameters were measured at WAND for the (Tl,Rb), Cs and Rb samples, and
at the fixed $E_i=14.7$ meV triple-axis spectrometer HB1A for the (Tl,K) one.
Sample temperature was regulated by high temperature Displex close cycle refrigerator.

\begin{figure*}
\includegraphics[width=175mm]{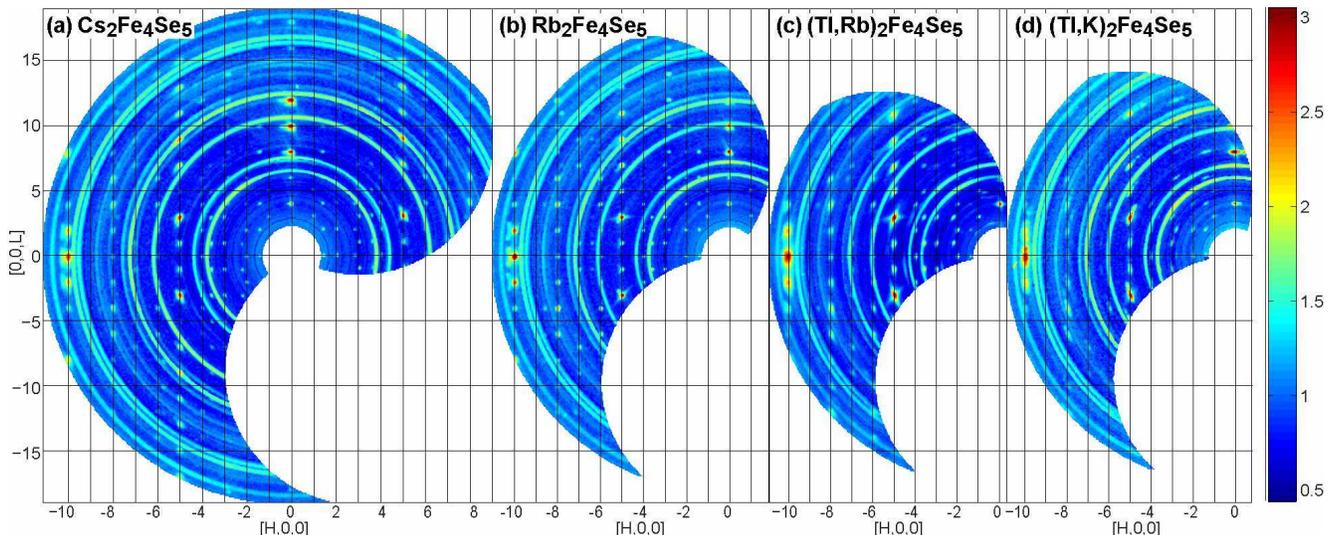}
\vskip -.4cm
\caption{Single crystal neutron diffraction pattern of (a) Cs$_{2}$Fe$_{4}$Se$_5$, (b) Rb$_{2}$Fe$_{4}$Se$_5$, (c) (Tl,Rb)$_2$Fe$_4$Se$_5$ and (d) (Tl,K)$_2$Fe$_4$Se$_5$ measured in the (h0$\ell$) plane of the $I4/m$ unit cell at T=295 K. In addition to Bragg peaks of the high temperature $I4/mmm$ structure, two new types of contributions to Bragg reflection appears: 1) structural \{200\}, \{202\},  \{400\}, \{402\}, etc.\ caused by the Fe vacancy order, and 2) magnetic \{103\}, \{301\}, \{303\}, \{701\}, etc.\ caused by 
antiferromagnetic order of the Fe moments.}
\label{fig2}
\end{figure*}

In term of the high temperature $I4/mmm$ unit cell of the BaFe$_2$As$_2$ structure, the Fe vacancy order is represented by the
appearance of the structural superlattice peaks characterized by ${\bf Q}_S=(\frac{1}{5},\frac{3}{5},0)$ \cite{D014882}, and the antiferromagnetic order by ${\bf Q}_M=(\frac{2}{5},\frac{1}{5},1)$ \cite{D020830}. Both types of superlattice peaks are accommodated by the $\sqrt{5}$$\times$$\sqrt{5}$$\times$1 tetragonal $I4/m$ unit cell \cite{D020830,D020488} which we use to label reciprocal space in this paper and in which ${\bf Q}_S=(110)$ and ${\bf Q}_M=(101)$.
Fig.~\ref{fig2} shows single-crystal diffraction pattern in the ($h0\ell$) plane at 295 K. Structural Bragg peaks \{h0$\ell$\} (h = 2, 4, 6, \dots\, and $\ell$ = 0, 2, 4, \dots) due to the Fe vacancy order as well as magnetic Bragg peaks \{h0$\ell$\} (h = 1, 3, 7, \dots and $\ell$ = 1, 3, 5, \dots) caused by the block checkerboard antiferromagnetic order \cite{D020830} are clearly visible. These ``superlattice'' peaks are forbidden in the BaFe$_2$As$_2$ structure. 

Like in previous x-ray and electron single crystal diffraction studies \cite{D014882,D020488,D012059}, twins with the same $c$-axis exist in all crystals examined in this neutron single crystal diffraction study. They show up as those ``superlattice'' peaks
at non-integer $h$ indices in Fig.~\ref{fig2}. For example, the $(-1.41,0,\ell)$ with $\ell$ even in Fig.~\ref{fig2}(a) can be identified as the Fe vacancy peaks ($11\ell$) from a twinned domain.
During the data reduction, care was paid to resolve the twin distribution. 
Data collected using WAND and FCD were consistent with each other. 
Data analysis was conducted using the FULLPROF program suite.
The relative occupancy of Tl and Rb (Tl and K) was fixed by that from the energy dispersive x-ray (EDX) spectroscopic measurements since our neutron diffraction refinements did not have enough sensibility for an independent determination. Fixing or floating the ratio does not affect appreciably the value of refined magnetic moment. The total number of independent Bragg reflections used in the refinements were 155, 131, 190 and 132 for the $A$=Rb, Cs, (Tl,K) and (Tl,Rb) samples, respectively, and the RF-factors 
were 4.8, 5.6, 4.5 and 7.7\%, respectively.
A summary of results of the four superconductors investigated in this study is listed in Table I, together with those of K$_{2}$Fe$_{4}$Se$_5$ from previous study \cite{D020830}.

\begin{table}[bt!]
\caption{Physical properties of the $A_{2}$Fe$_{4}$Se$_5$ superconductors.
Lattice parameters at 295 K, ordered magnetic moment of Fe at 295 K ($m$)
and at $\sim$10 K ($M$) are provided together with $T_c$, $T_N$ and Fe vacancy order-disorder transition $T_S$.
}
\label{tab1}
\begin{tabular}{l|ccccc}
\hline \hline
 $A$ & K & Rb  & Cs  & Tl,K & Tl,Rb \\
    \hline
  a ($\AA$) & 8.7306(1) & 8.788(5) & 8.865(5) & 8.645(6) & 8.683(5)  \\
 c ($\AA$) & 14.1128(4)& 14.597(2) & 15.289(3) & 14.061(3)& 14.388(5)  \\
  $ m$ ($\mu_B$)& 2.76(8)  & 2.95(9) & 2.9(1)& 2.61(9)&2.7(1)   \\
  $ M$ ($\mu_B$)& 3.31(2)  & 3.3(1)  & 3.4(2)& 3.2(1)  & 3.2(1)   \\
 $T_c$ (K) & 32 & 32 & 29 & 28 & 32 \\
 $T_N$ (K) & 559(2) & 502(2) & 471(4) & 506(1)  &  511(1) \\
$T_S$ (K) & 578(2) & 515(2) & 500(1) & 533(2)  & 512(4) \\    
\hline \hline
    \end{tabular}
\end{table}

\begin{figure}[tb!]
\includegraphics[width=.9\columnwidth]{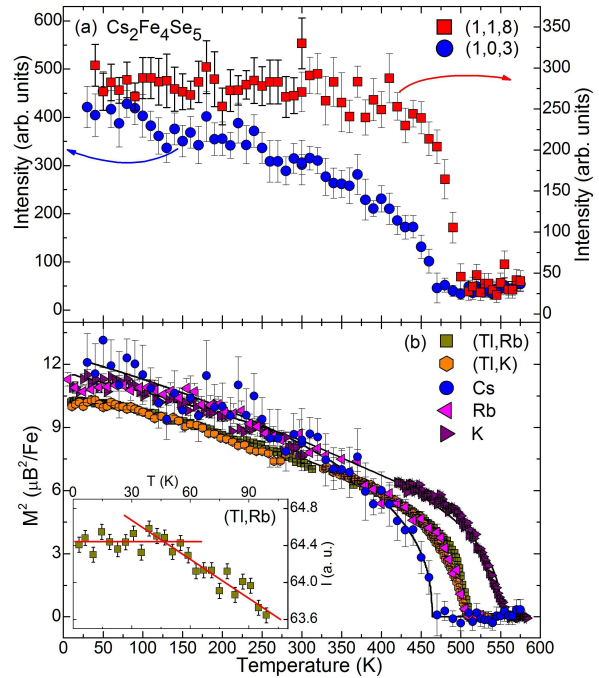}
\vskip -.3cm
\caption{(a) Magnetic (103) and structural (118) Bragg peaks vs.\ temperature, serving as order parameters for the antiferromagnetic and Fe vacancy order-disorder transitions, respectively, in Cs$_{2}$Fe$_{4}$Se$_5$.  
(b) Normalized magnetic Bragg intensity representing the squared magnetic moment as a function of temperature for $A_{2}$Fe$_{4}$Se$_5$ superconductors. Inset: Magnetic (101) peak of (Tl,Rb)$_{2}$Fe$_{4}$Se$_5$ showing that its intensity saturates when $T_c$ is approached.
}
\label{fig3}
\end{figure}

Two separated phase transitions due to antiferromagnetic order and the iron vacancy order have been discovered in neutron diffraction study on K$_{2}$Fe$_{4}$Se$_5$ \cite{D020830}, but only magnetic transition has been reported previously for the Cs compound in a combined $\mu$SR and thermometry study \cite{D011873}. In Fig.~\ref{fig3}(a), integrated intensity of magnetic Bragg peak (103) and structural Bragg peak (118) of Cs$_{2}$Fe$_{4}$Se$_5$, simultaneously measured on WAND, are shown as a function of temperature. Similar to the case of the K intercalated superconductor, there are two separated magnetic and structural transitions at $T_N= 471(4)$ K and $T_S = 500(1)$ K, respectively.  
The order parameters at ${\bf Q}_S$ and ${\bf Q}_M$ look very
similar to those of K$_{2}$Fe$_{4}$Se$_5$ \cite{D020830}, albeit now
that magnetic and structural transition temperatures of $T_N$ and $T_S$ are lower [Table I and Fig.~\ref{fig3}].
The $T_N$ reported here agrees with $T_N\approx 477$ K from the previous $\mu$SR study \cite{D011873},
which also reports a microscopic coexistence of superconductivity and an unidentified antiferromagnetic order in a nominal Cs$_{0.8}$Fe$_2$Se$_{1.96}$ ($T_c\approx 28.5$ K) superconductor. The $T_S$ agrees with that from an x-ray study appearing after we finished the measurement \cite{D021919}. 
Different from previous Fe-based superconductors where antiferromagnetic order coexisting with superconductivity can be regarded as relic from the parent antiferromagnet \cite{A073950}, antiferromagnetic order in 245 superconductors is very strong. In addition to the high $T_N$, all the magnetic moments listed in Table I at $\sim$10 K are higher than previous record
2 $\mu_B$/Fe found in nonsuperconducting {\em parent} compounds \cite{A092058}.

The structural peak at ${\bf Q}_S$, which represents the breaking of the high temperature $I4/mmm$ symmetry when the vacant (orange cube in Fig.~\ref{fig1}) and occupied (purple sphere in Fig.~\ref{fig1}) Fe sites segregate below $T_S$, saturates rapidly to its low temperature value [Fig.~\ref{fig3}(a)]. Therefore, below the room temperature where relevant physics processes relating to the superconductivity occur, we are dealing with a well ordered crystal structure in the $I4/m$ symmetry. Structurally, and consequently electronically, using the BaFe$_2$As$_2$ family of superconductor materials as a model is too remote to the reality. The modification of electronic structure by the new $I4/m$ crystal symmetry and by the block antiferromagnetic order has been vividly and dramatically demonstrated in first-principle band structure calculations \cite{D021344,D022215}.
Future investigations on the 245 superconductors should be based on the correct $I4/m$ crystal structure.

Comparing to the temperature dependence of the structural peak, the magnetic
Bragg peak of Cs$_{2}$Fe$_{4}$Se$_5$ show a gradual rise below $T_N$
[Fig.~\ref{fig3}(a)], as previously found for K$_{2}$Fe$_{4}$Se$_5$
\cite{D020830}.  The $T_N$ is lower than $T_S$ as expected, since the
long-range magnetic order builds on the ordered Fe pattern.  This is
reminiscent of the antiferromagnetic transition in V$_2$O$_3$ albeit the
underlying transition is to an orbital occupation order \cite{bao96c}.
Fig.~3(b) illustrates the magnetic order parameters of all five $\rm A_2Fe_2Se_5$
samples with $T_N$s listed on Table I. Like in the K compound, the linear
dependence of magnetic Bragg intensity is interrupted by a flat plateau when
$T_c$ is approached. For example, the inset of Fig.~\ref{fig3}(b) shows a
temperature scan measured with much better statistics using triple-axis
spectrometer for the $\rm (Tl,Rb)_2Fe_4Se_5$ sample, suggesting interaction
between the antiferromagnetic order and superconductivity as previously
discovered in the case of heavy fermion superconductors \cite{UPt3_GA}.

Since the discovery of the nominal K$_{0.8}$Fe$_2$Se$_2$ superconductor \cite{C122924}, 
there had been chaotic confusion about sample composition of the new iron selenide superconductors due to the lack of reliable structure determination work. The Fe-Se square lattice was initially regarded as intact as in the previous families of iron-based superconductors.
Analogue to the heavily electron doped iron pnictide superconductors was advocated.
Theories based on both the $A$Fe$_{2}$Se$_2$ of the BaFe$_2$As$_2$ structure and $A$Fe$_{1.5}$Se$_2$ of a $\sqrt{2}$$\times$$2\sqrt{2}$$\times$1 supercell structure framework were advanced for the new superconductors \cite{D014390,C126015}. The superconductivity in the (Tl,K) system appeared to occur at a magnetic quantum critical point \cite{C125236} and supported a doped Mott insulator scenario \cite{D010533,D013307} similar to the case of cuprate superconductors.
While electronic properties are indeed very different from previous iron-based superconductors as
probed in ARPES \cite{C125980,D014556,D014923}, optic \cite{D010572},
NMR \cite{D011017,D014572,D014967}, $\mu$SR \cite{D011873} and Raman scattering \cite{D012168} studies, an appropriate understanding cannot be achieved without knowing the superconducting composition, crystal structure and magnetic order determined in neutron and x-ray diffraction structural studies.

In summary, our systematic neutron diffraction works show that Cs,
Rb, (Tl,Rb) and (Tl,K) intercalated Fe selenide superconductors are
members of the common family of the $A_{2}$Fe$_{4}$Se$_5$
superconductors which share the {\it same} Fe vacancy- and antiferromagnetic order. Not only the crystal structure of
the $A_{2}$Fe$_{4}$Se$_5$ superconductors in the $\sqrt{5}$$\times$$\sqrt{5}$$\times$1 $I4/m$ unit cell is very different from that in previous iron-based high-$T_c$ superconductors, leading to drastically different Fermi surface topology,
the coexistence of superconductivity with the very high $T_N$ and large moment antiferromagnetic order is unprecedented. Although the same Fe ions are involved, given the drastically different crystalline and magnetic structures, the $A_{2}$Fe$_{4}$Se$_5$ superconductors are unlikely to share the same
physics mechanism as in previously discovered families of iron-based superconductors.
Future search for new high-$T_c$ superconductors has a wider playground including vacancy decorated
lattice.

We thank Q. Huang and M. A. Green for useful discussions. 
The works at RUC, USTC and ZU were supported by the NSFC Grants 11034012, 10974175 and 10934005 and by the 973 Program Grants 2011CBA00112, 2011CBA00103 and 2009CB929104. The work at ORNL was  supported by the Division of Scientific User Facilities, DOE OBES. 
WAND is operated jointly by ORNL and JAEA under the US-Japan Cooperative Program in Neutron Scattering.


\end{document}